\documentclass[11pt]{article}
\def\la{\langle}
\def\ra{\rangle}

\def\ra{\rangle}
\usepackage{epsfig}
\begin{document}
\begin{titlepage}

\vspace{4cm}

\begin{center}{\Large \bf Thermal entanglement of spins in an inhomogeneous magnetic field}\\
\vspace{1cm} M. Asoudeh\footnote{email:asoudeh@mehr.sharif.edu},
\hspace{0.5cm} V. Karimipour \footnote{Corresponding author, email:vahid@sharif.edu}\\
\vspace{1cm} Department of Physics, Sharif University of Technology,\\
P.O. Box 11365-9161,\\ Tehran, Iran
\end{center}
\vskip 3cm

\begin{abstract}
We study the effect of inhomogeneities in the magnetic field on
the thermal entanglement of a two spin system. We show that in the
ferromagnetic case a very small inhomogeneity is capable to
produce large values of thermal entanglement. This shows that the
absence of entanglement in the ferromagnetic Heisenberg system is
highly unstable against inhomogeneoity of magnetic fields which
is inevitably present in any solid state realization of qubits.
\end{abstract}
\end{titlepage}

\section{Introduction}\label{intro}
\subsection{Motivation}\label{motiv} It is well known that quantum entanglement \cite{epr,
sch, bell} plays a fundamental role in almost all  efficient
protocols of quantum computation (QC) and quantum information
processing \cite{bennett, chuang}. \\
Without entanglement which is the essential quantum ingredient of
QC, any quantum algorithm which only uses the other property of
quantum mechanics, namely the superposition property, can also be
implemented on any physical system which allows superposition of
states, i.e. classical linear optical devices. In any proposal
for physical implementation of qubits, it is therefore of utmost
importance to investigate the entanglement properties of pairs
and collection of such qubits. Among the many proposals for
physical implementation of qubits, those based on solid state
devices seem to be promising as far as the crucial scalablity
property is concerned. \\ In one such proposal \cite{kane} a well
localized nuclear spin coupled with an electron of a donor atom
in silicon plays the role of a qubit which can be individually
initialized, manipulated and read out by extremely sensitive
devices. In another proposal\cite{loss, divin1, burkard,
imamoglu}, the spin of an electron in a quantum dot plays the
role of a qubit. Long decoherence time and scalability to more
than 100 qubits are two of the
important virtues of this scheme. \\
In both schemes the effective interaction between the two qubits
is governed by an isotropic Heisenberg Hamiltonian with Zeeman
coupling of the individual spins, namely
\begin{equation}\label{eff}
  H = J {\bf S}_1\cdot {\bf S}_2 + \gamma (S_{1z} +
  S_{2z}).
\end{equation}
Actually the isptropic interaction is an approximation, since
spin orbit coupling introduce perturbations which break this
isotropy . A more complete hamiltonian would be \cite{kavokin, wu}
\begin{equation}\label{}
H = J({\bf S}_1\cdot {\bf S}_2 + \overrightarrow{\beta}\cdot {\bf
S}_1\times {\bf S}_2  +\gamma \overrightarrow{\beta}\cdot {\bf
S}_1\overrightarrow{\beta}\cdot {\bf S}_2) +\gamma (S_{1z} +
  S_{2z}),
\end{equation}
where the dimensionless vector $\overrightarrow{\beta}$ is called
the spin orbit field and in systems like the GaAs quantum dots has
a magnitude $|\overrightarrow{\beta}|$ of a few percent and the
dimensionless $\gamma$ is of the order of $10^{-4}$. Note that the
only coupling in the interaction between spins that is
controllable is $J$ \cite{wu}, and the individual couplings
between different components of spins denoted usually by $J_x,
J_y$ and $J_z$ can not be controlled separately and thus one can
not adjust these parameters arbitrarily to enhance the
entanglement in a given situation.\\ This means that although
studies of entanglement for different types of anisotropic
interactions are very interesting theoretically (specially when
infinite spin systems are treated which is the only case which
yields valid results with regard to quantum phase transitions
\cite{sachdev}), they may not be of much practical relevance to
concrete physical realization of qubits.\\  In this paper we
ignore the anisotropic perturbations due to both their smallness
and due to the fact
that strategies have been invented to cancel such anisotropies \cite{bonesteel}.\\
Due to their smallness, they may only introduce minor changes in
any result derived for the
isotropic case.\\
On the other hand in any solid state construction of qubits,
there is always the possibility of inhomogenous Zeeman coupling
\cite{Hu1, Hu2}. Solid state heterostructures are usually
inhomogeneous and magnetic imperfections or impurities are likely
to be present leading to stray magnetic fields. Indeed it is one
of the main challenges in this proposal to construct identical
qubits \cite{ibm}. Constructing nearly identical devices in
semiconductor technology has always been difficult and is still
difficult, e.g. a very small temperature or strain difference in
the substrate produces differences which although may not be
significant for the classical semiconductor technology will
certainly be important for the quantum technology \cite{ibm}.
Besides these unwanted effects, there are schemes like parallel
pulsed schemes \cite{pps}, in which both a localized and hence
inhomogenous Zeeman coupling and exchange interactions are
employed to expedite manipulation of qubits.\\ \\
In view of the above it is desirable to consider a two qubit
system in an inhomogenous magnetic field and study the
entanglement properties
of this system in detail. \\
At extremely low temperatures such a qubit system may be assumed
to be in its ground state. Thus it will be desirable to study the
entanglement properties of the ground state. \\
However a real physical system is always at a finite temperature
and hence in a mixture of disentangled and entangled states
depending on the temperature. Therefore one is naturally led to
consider the thermal entanglement of such physical systems.\\ In
summary we mean that the thermal entanglement of finite systems
has more relevance to the problem of initialization of quantum
computers \cite{divin2} than to the problem of quantum phase
transitions which requires a study of infinite size systems.\\
\subsection{A brief account of previous works}  Thermal entanglement in a two qubit Heisenberg
magnet with the Hamiltonian
\begin{equation}\label{ham}
    H=J\overrightarrow{\sigma}_1\cdot \overrightarrow{\sigma}_2 +
    B(\sigma_{1z} +\sigma_{2z}).
\end{equation}
was first studied by Nielsen \cite{nielson} who showed that in
the ferromagnetic case ($J<0$) no entanglement exists but in the
antiferromagnetic case ($J>0$) entanglement appears below a
threshold temperature $T_c$. Since then many other systems have
been investigated.\\ There is now a vast literature on this
subject and for clarity it is better to separate them into two
categories, namely those \cite{oconner, arnesen, wz, osterloh,
osborne} which study by analytical or numerical methods infinite
spin chains with at times particular attention to quantum phase
transitions and those which study few, mostly two, spin systems.
In our opinion one can not draw valid results for quantum phase
transitions by studying a two spin system, and these types of
studies are useful in other contexts, e.g. the problem of
initialization of a quantum computer as described above, provided
they start with a plausible hamiltonian
for the interaction of physical qubits.\\
In the following we mention some of the works only in this latter
category which are of relevance to our work in this paper. \\
After the work of Nielson \cite{nielson}, it has been shown that
two spins interacting by the Ising interaction in the $z$
direction, when placed in a magnetic field of arbitrary
direction, acquire maximum entanglement when the magnetic field
is perpendicular to
the $z$ direction \cite{gunlycke}.\\
The effect of anisotropy (in the spin couplings in the $x$, $y$
and $z$ directions) has also been studied in a number of works for
different models \cite{rigolin, wang, kamta, canosa}. The effect
of inhomogeneous magnetic fields has been studied in \cite{sun},
but only on an $XY$ system. Such a system already shows
entanglement when placed in a uniform magnetic field.
\subsection{Results}
In this paper we have studied an isotropic two qubit system in a
inhomogeneous magnetic field, described by the Hamiltonian

\begin{equation}\label{ham2}
    H=J\overrightarrow{\sigma}_1\cdot \overrightarrow{\sigma}_2 +
    (B+b)\sigma_{1z} + (B-b)\sigma_{2z},
\end{equation}
where $J$ is the isotropic coupling between the spins, $B \geq 0
$, and the magnetic fields on the two spins have been so
parameterized that $b$ controls the degree of inhomogeneity.\\ \\
Let us first review the situation for the homogeneous magnetic
field.\\

For the ferromagnetic ($J<0$) system, there is no thermal
entanglement at any temperature, but for the anti-ferromagnetic
($J>0$) case, thermal entanglement develops when the temperature
drops below the threshold value $ kT_c:= \frac{4J}{\ln 3}$. We
want to see how the presence of inhomogeneity modifies this
situation. We will show that inhomogeneity has the following
effects: \\

1- In the ferromagnetic system it generally produces entanglement,
dependent on the value of the magnetic field and the temperature.
There is a threshold temperature above which no entanglement is
possible. This temperature has in fact been zero in the uniform
case which has been shifted to finite values by the
inhomogeneity. Specially at temperatures near zero and in zero
magnetic field, the effect of inhomogeneity is very significant.
Under this condition a very small inhomogeneity produces maximal
entanglement as shown in figures (3 and 4).
\\

2-  In contrast to the ferromagnetic case, the effects in the
anti-ferromagnetic system are small. Inhomogeneity in this case
slightly raises the threshold temperature, and lowers the value of
entanglement as shown in figures (5 and 6).
\\ \\ The structure of this paper is as follows: After presenting the essentials of thermal entanglement
in the next section, in section \ref{ground}
 we study the spectrum of the Hamiltonian and characterize
the entanglement of the ground state in various regions of the
parameter space. In section \ref{thm} we analyze the thermal
entanglement of the system. Throughout the paper we normalize the
coupling between spins to $J=1$ for the anti-ferromagnetic case
and to $J=-1$ for the ferromagnetic case and study the results for
the two cases separately.

\section{Preliminaries on thermal
entanglement} A spin system with Hamiltonian $H$ kept at
temperature $T$ is characterized by a density matrix
$\rho:=\frac{1}{Z}e^{-\beta H}$, where $\beta = \frac{1}{kT}$, $k$
is the Boltzman constant and
$Z:=tr e^{-\beta H}$ is the partition function.\\
The entanglement of this density matrix, called the thermal
entanglement of the spin system can be calculated exactly with the
help of Wootters formula \cite{wootters}. Explicitly it is given
by the following formula

\begin{equation}\label{E}
    E(\rho) = -\frac{1+\sqrt{1-C^2}}{2}\log_2 {\frac{1+\sqrt{1-C^2}}{2}}
    - \frac{1-\sqrt{1-C^2}}{2}\log_2 {\frac{1-\sqrt{1-C^2}}{2}},
\end{equation}
where
\begin{equation}\label{Clambda}
  C = max \{ 0, \lambda_1-\lambda_2-\lambda_3-\lambda_4\},
\end{equation}
and  $\lambda$'s  are the positive square roots of the eigenvalues
of the matrix $\rho\tilde{\rho}$ in decreasing order. The matrix
$\tilde{\rho}$ is defined as
\begin{equation}\label{rho'}
\tilde{\rho} = (\sigma^y \otimes \sigma^y)\rho^*(\sigma^y \otimes
\sigma^y),
\end{equation}
where $*$ denotes complex conjugation in the computational
basis.\\ In case that the state is pure $\rho = |\psi\ra\la
\psi|$, with
\begin{equation}\label{pure} |\psi\ra:= a|+,+\ra +
b|+,-\ra + c|-,+\ra + d|-,-\ra
\end{equation}
the above formula for the concurrence is simplified to
\begin{equation}\label{Cpure}
C(\psi)=2|ad-bc|.
\end{equation}
Since $E$ is an increasing function of $C$, it is usual to take
$C$ itself as a measure of entanglement whose value ranges from
$0$ for a disentangled state to $1$ for a maximally entangled
state. In the following sections we apply this formalism to the
inhomogeneous system given by the Hamiltonian (\ref{ham2}).

\section{Ground state entanglement}\label{ground}
When the magnetic field is uniform, i.e. $b=0$, the hamiltonian
(\ref{ham2}) has two symmetries, namely $[H,S_z]=[H,S^2]=0$, where
$S_z$ and $S^2$ are the third component of spin and the total spin
respectively. In a inhomogeneous magnetic field, the symmetry
$[H,S^2]=0$ no longer holds and thus the triplet and the singlet
spins are no longer energy eigenstates separately. A
straightforward calculation gives the following eigenstates:
\begin{eqnarray}\label{states}
|\phi_1\ra &=&|+,+\ra  \cr |\phi_2\ra &=&|-,-\ra \cr |\phi_3\ra
&=&
\frac{1}{\sqrt{2({\delta}^2+(1-\xi)^2)}}\left((\delta-1+\xi)|+,-\ra+(\delta+1-\xi)|-,+\ra\right)\cr
 |\phi_4\ra &=&
\frac{1}{\sqrt{2({\delta}^2+(1-\xi)^2)}}\left((\delta+1-\xi)|+,-\ra-(\delta-1+\xi)|-,+\ra\right),
\end{eqnarray}
with corresponding energies
\begin{eqnarray} E_1 &=& J+2B\cr E_2 &=& J-2B \cr E_3&=& -J(1-2\xi)\cr  \ E_4 &=& -J(1+2\xi),
\end{eqnarray}
where $\xi:=\sqrt{1+\delta^2}$ and $\delta = \frac{b}{J}$.\\
Note that we are working in units so that $B$ and $J$ are
dimensionless. It turns out that $\xi$ is the suitable parameter
for expressing the effects of inhomogeneity. Thus hereafter we
will mostly use $\xi$ rather than the original parameter $b$, in
our analysis. The value $\xi = 1 $ corresponds to a uniform
magnetic field and deviations from this value characterize the
degree of non-uniformity.
 \\In
the limiting case $\xi\longrightarrow 1 $, the two states
$|\phi_1\ra $ and $|\phi_2\ra$ respectively go to the maximally
entangled states $\frac{1}{\sqrt{2}}(|+,-\ra + |-,+\ra)$ and
$\frac{1}{\sqrt{2}}(|+,-\ra - |-,+\ra)$.
\subsection{The ferromagnetic case, $\ J=-1$}
The ground state depends on the value of the magnetic field $B$
and the inhomogeneity parameter $\xi$. It is readily found that
the ground state energy is equal to:
\begin{eqnarray}
  \left\lbrace
  \begin{array}{l}
    E_2=-1-2B  \ \ \ \ \ {if\ \ \ \ \ \xi< B+1},\\
\\
   E_3=\  1-2\xi  \ \ \ \ \ \ {if \ \ \ \ \ \xi> B+1}
  \end{array}\right.
\end{eqnarray}

\begin{figure}\label{mine0}
\begin{center}
\epsfig{file=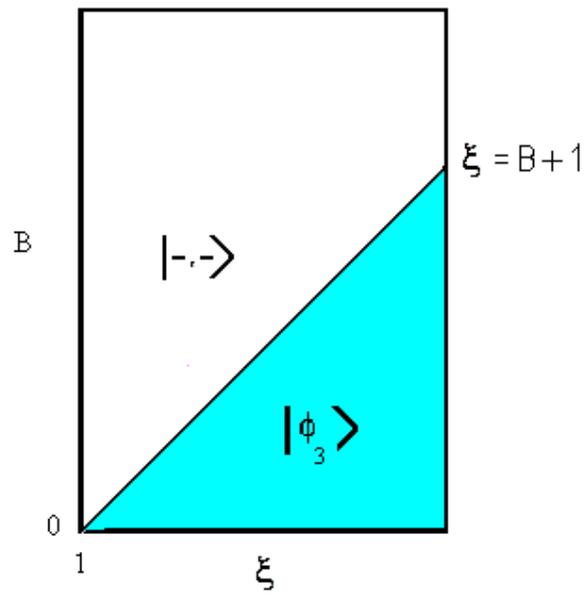,width=10cm} \caption{(color online)The
ground state of the ferromagnetic case, as a function of
inhomogeneity $\xi$ and the magnetic field $B$. We work in units
where $B$ is dimensionless.}
\end{center}
\end{figure}

Thus for $\xi < B+1 $, the ground state is the disentangled state
$|\phi_2\ra$  and for $\xi> B+1$,  the ground state is the
entangled state $|\phi_3\ra$. \\ The phase diagram of the ground
state is shown in figure (1).\\
For each value of the magnetic field $B$, there is a threshold
parameter $\xi^f:=B+1 $ above which the ground state will become
entangled. Conversely for each value of inhomogeneity $\xi$ there
is a value of magnetic field $B^f:=\xi-1$ above
which the ground state will loose its entanglement.\\
In the entangled phase the entanglemenet of the ground state is
found from (\ref{Cpure}) and (\ref{states}) to be

\begin{equation}\label{Cphi3}
C(\phi_3)=\frac{1}{\xi},
\end{equation}

which is solely determined by inhomogeneity. A very interesting
point is that when $B=0$, with an infinitesimal value of $b\approx
0$ ($\xi\approx 1$) the system  enters the maximally entangled
phase $|\phi_3\ra$ with etanglment $C=\frac{1}{\xi}\approx 1$.
This remarkable feature means that the absence of entanglement in
ferromagnetic Heisenberg chain is completely unstable against very
small inhomogeneities. It is also reminiscent of quantum phase
transitions where a slight change in one of the parameters of the
system, changes the behavior of the system dramatically.
Increasing further the inhomogeneity will move the ground state
further into the entangled phase but reduces its entanglement due
to (\ref{Cphi3}).

\subsection{The anti-ferromagnetic case, $\ J=1$}
In this case we find that the ground state energy is equal to

\begin{eqnarray}
 \left\lbrace
  \begin{array}{l}
    E_2=1-2B  \ \ \ \ \ \ {if\ \ \ \ \xi<  B-1},\\
\\
   E_4=\  -1-2\xi  \ \ \ \ \ \ {if \ \ \ \ \xi> B-1}
  \end{array}\right.
\end{eqnarray}

\begin{figure}\label{mine00}
\begin{center}
\epsfig{file=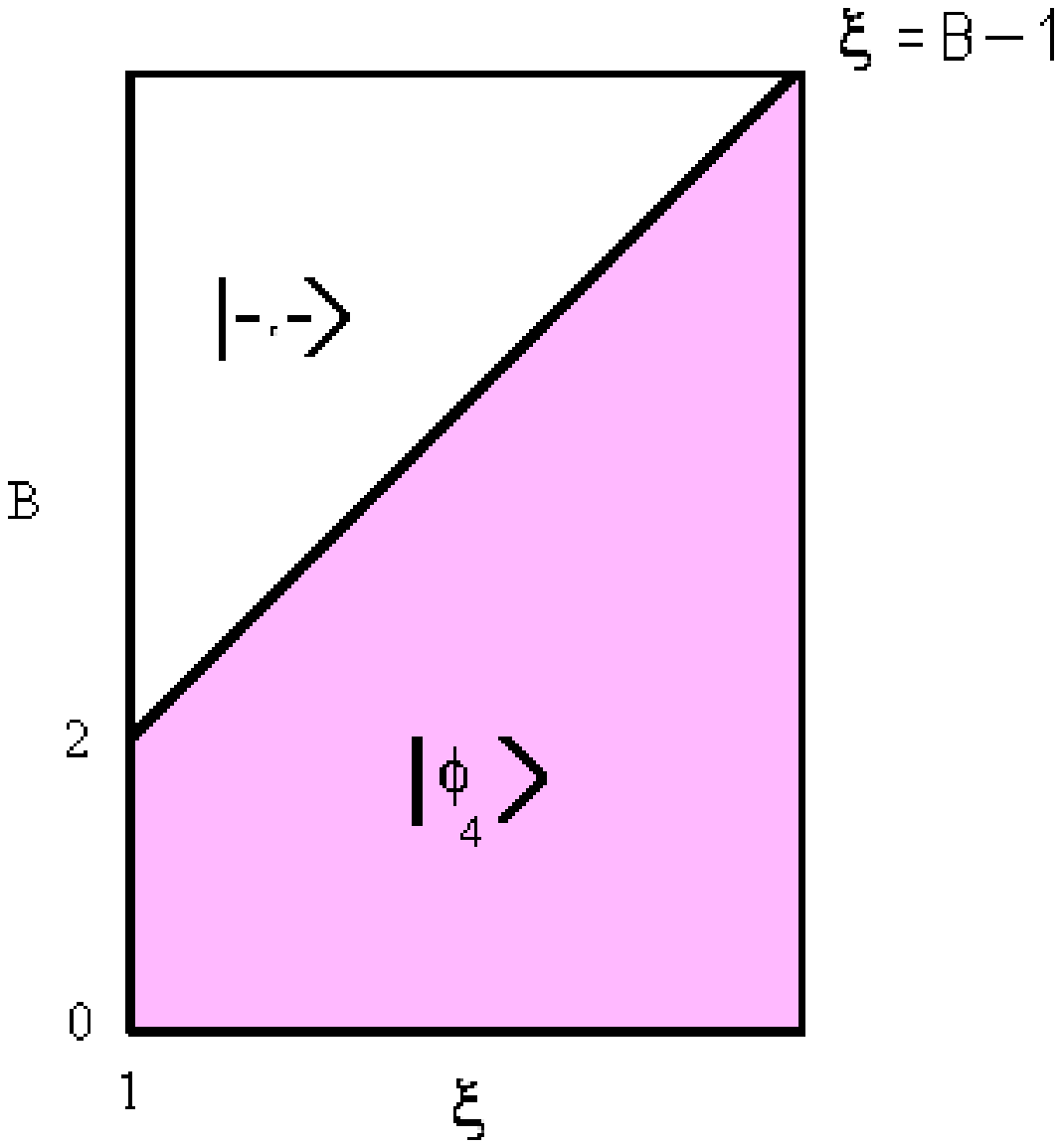,width=10cm} \caption{(color online) The
ground state of the antiferrmagnetic case, as a function of
inhomogeneity $\xi$ and the magnetic field $B$. We work in units
where $B$ is dimensionless.}
\end{center}
\end{figure}

Thus for  $\xi < B-1 $, the ground state is the disentangled state
$|\phi_2\ra$ and for $\xi> B-1$, the ground state is the entangled
state $|\phi_4\ra$. The phase diagram of the ground state is shown
in figure (2).\\

Again in the entangled phase the entanglement of the ground state
is found from (\ref{Cpure}) and (\ref{states}) to be
\begin{equation}\label{Cphi4}
    C(\phi_4)= \frac{1}{\xi}.
\end{equation}
which is independent of $B$. Increasing inhomogeneity again
decreases
the concurrence and hence the entanglement.\\

\section{Thermal entanglement}\label{thm}
Raising the temperature mixes the ground state with excited
states. Depending on the sign of $J$ and the value of parameters
this may increase or decrease the value of entanglement. In some
cases the disentangled ground state mixes with entangled excited
states and in some other cases the entangled ground state mixes
with disentangled excited states. To see what happens exactly we
calculate the entanglement of the thermal state $\rho =
\frac{1}{Z} e^{-\beta H}$. The symmetry $[H,S_z]=0$ constrains the
general form of $\rho $ to

\begin{equation}\label{rho}
    \rho=\left(\begin{array}{cccc}
      u_+ &  &  &  \\
       & w & z &  \\
       & z & w &  \\
       &  &  & u^- \\
    \end{array}\right),
\end{equation}
where  $C$ is found from (\ref{Clambda} and \ref{rho'}) to be
given \cite{oconner} by
\begin{equation}\label{C}
  C = 2 \ {\rm {max}}\ (0, |z|-\sqrt{u^+u^-}).
\end{equation}
The exact values of the elements of $\rho $ is obtained by knowing
the spectrum of $H$. After a simple calculation from
\begin{equation}\label{rhoexpansion}
    \rho =\frac{1}{Z}\sum_{i=1}^{4} e^{-\beta E_i}|\phi_i\ra\la \phi_i|
\end{equation}
we obtain
\begin{eqnarray}\label{uu}
    u_+&=&\frac{1}{Z}e^{-\beta(J+2B)},\cr
    u_-&=&\frac{1}{Z}e^{-\beta(J-2B)},
\end{eqnarray}
and
\begin{equation}
    z =
    \frac{-1}{Z}\frac{1}{\xi}e^{\beta J}\sinh{2J\beta\xi},
\end{equation}
where $Z$ is the partition function given by
\begin{equation}\label{Z}
    Z:= tr e^{-\beta H}=2e^{-\beta J} \cosh{2\beta B} + 2e^{\beta J}\cosh{2\beta
    J\xi}.
\end{equation}
Thus from (\ref{C}) we find that
\begin{equation}\label{Cfinalgeneral}
    C=\frac{2}{Z}\ \ {\rm{Max}}\ \left(0\  ,\  \frac{1}{\xi}e^{\beta
    J}|\sinh{2\beta J\xi}|-e^{-\beta J}\right).
\end{equation}
We consider the ferromagnetic ($J=-1$) and the anti-ferromagnetic
($J=1$) cases separately.
\subsection{Ferromagnetic case, $ J=-1$}
Setting $J=-1$ in (\ref{Cfinalgeneral}) we have
\begin{equation}\label{Cfinalferro}
    C = {\rm{Max}}\ \left(0,\frac{e^{-\beta} \sinh{2\beta \xi}-\xi e^{\beta}}{\xi(e^{\beta}\cosh {2\beta B}
     + e^{-\beta}\cosh {2\beta\xi})}
\right).
\end{equation}
The threshold temperature is obtained from the equation
\begin{equation}\label{thresholdf}
e^{-2\beta}\sinh{2\beta \xi}= \xi.
\end{equation}
In the uniform case ($\xi=1$), this equation turns into
$e^{4\beta} = -1$ which has no solution. Thus in this limit there
is no thermal entanglement in the spin system in accordance with previous results (\cite{nielson, arnesen, wz}).\\
However in the inhomogeneous case ($\xi\ne 1$) this equation has
nontrivial solutions. Figure (7) shows the variation of threshold
temperature
with $\xi$.\\
\begin{figure}\label{mar1}
\begin{center}
\epsfig{file=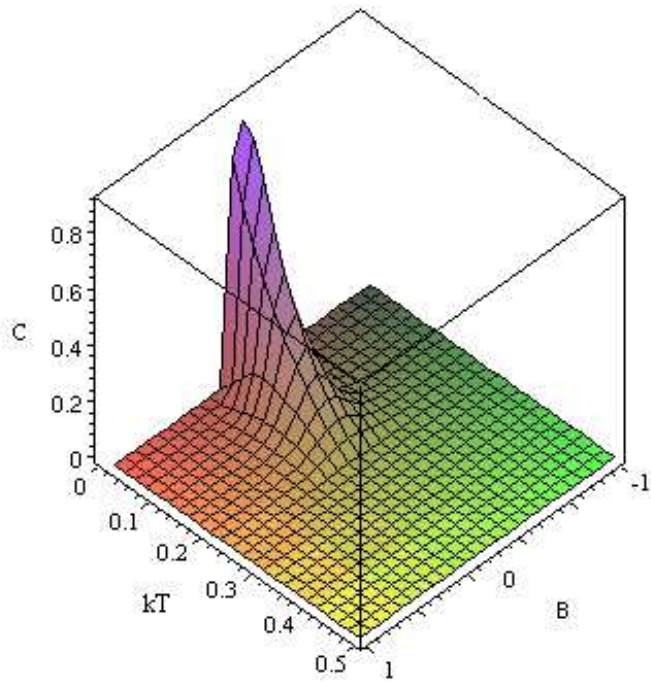, width=10cm}\caption{(color online)
Concurrence versus temperature and magnetic field for $\xi = 1.1$
in the ferromagnetic system.}\end{center}
\end{figure}
Figure (3) shows the entanglement as measured by the concurrence
for a fixed value of inhomogeneity $\xi=1.1$ in terms of the
temperature and magnetic field. Below the threshold temperature
(about $0.25$ for this value of $\xi$), thermal entanglement
develops and is maximized for zero magnetic field $B$. The value
of this maximum entanglement occurs of course at $T=0$, where its
value is equal to $\frac{1}{\xi}$, equal to 0.9 in this case.\\ \\
Figure (4) shows the value of entanglement in terms of the
temperature and the inhomogeneity for zero magnetic field.
\begin{figure}\label{mar2}
\begin{center}
\epsfig{file=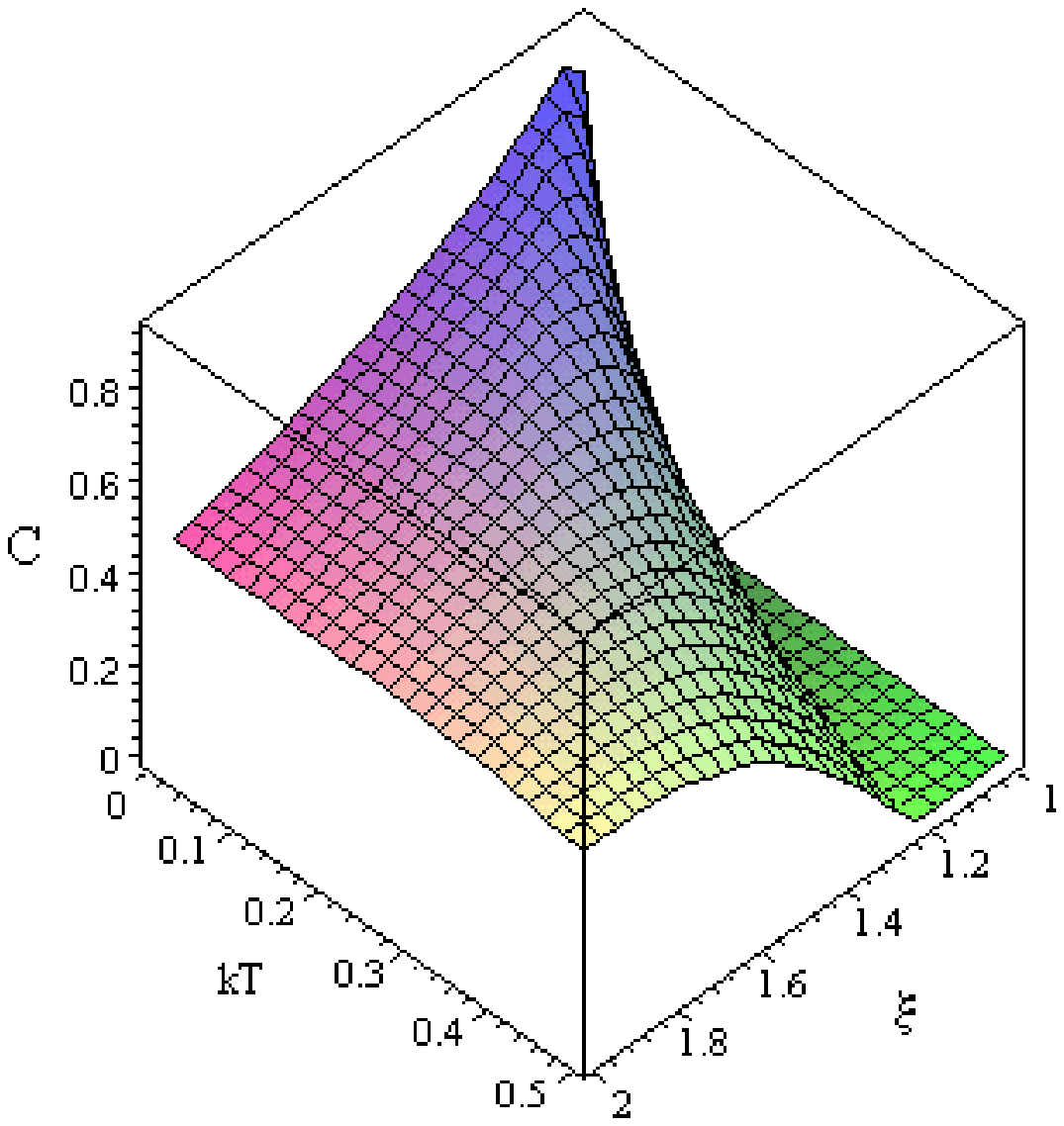,width=10cm }\caption{(color online)
Concurrence versus temperature and inhomogeneity in zero magnetic
field in the ferromagnetic system.}\end{center}
\end{figure}
 It is seen that at any temperature there is a parameter $\xi_0$
 above which thermal entanglement will develop in the system. The
 value of $\xi_0$ is obtained from (\ref{thresholdf}) and
 increases with increasing the temperature. At very low temperatures $\xi_0$ is very close to $1$ which shows that
 a small degree of inhomogeneity will develop maximal entanglement in the system.\\
\subsection{Anti-Ferromagnetic case} Setting $J=1$ in
(\ref{Cfinalgeneral}) we obtain
\begin{equation}\label{Cfinalantiferro}
  C = {\rm{Max}}\ \left(0,\frac{e^{\beta} \sinh{2\beta \xi}-
  \xi e^{-\beta}}{\xi(e^{-\beta}\cosh {2\beta B} + e^{\beta}\cosh {2\beta\xi})}\right)
\end{equation}

The threshold temperature is obtained from the equation
\begin{equation}\label{thresholdaf}
e^{2\beta}\sinh{2\beta \xi}= \xi.
\end{equation}
In the uniform case ($\xi=1$), this equation turns into
$e^{4\beta} = 3$ which gives the threshold temperature $kT_c =
\frac{4}{\ln 3}$ .  \\
In the inhomogeneous case ($\xi\geq 1$) this equation can be
solved numerically, the results is shown in figure (7). It is
seen that inhomogeneity only slightly increases the threshold
temperature in contrast to the ferromagnetic case where it had
appreciable effect.\\
Figure (5) shows the entanglement as measured by the concurrence
for a fixed value of inhomogeneity $\xi=1.1$ in terms of the
temperature and the magnetic field and figure (6) shows the value
of entanglement in terms of the temperature and the inhomogeneity
for zero magnetic field. Comparing these figures with figure (3)
and with the corresponding figure of (\cite{arnesen}) we see that
in the anti-ferromagnetic case, inhomogeneity has a small effect
on the threshold temperature and magnetic field and only
decreases the value of entanglement once it is developed. Its
value is weakened by raising the temperature and near the
threshold temperature it has a vanishingly small effect.
\begin{figure}\label{mar3}
\begin{center}
\epsfig{file=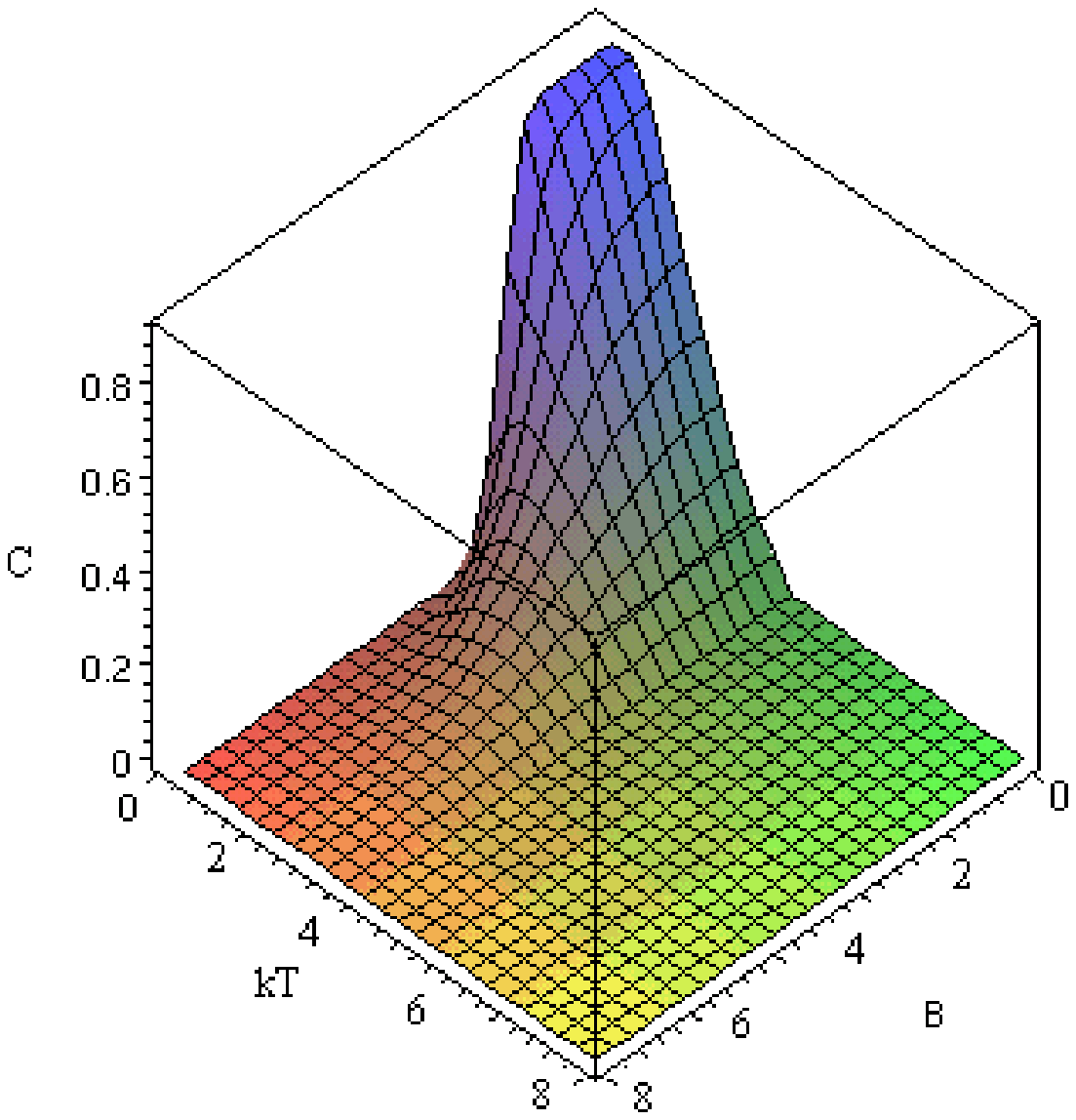,width=10cm}\caption{(color online)
Concurrence versus temperature and magnetic field for $\xi = 1.1$
in the anti-ferromgnetic system.}\end{center}
\end{figure}
It is seen that for any fixed temperature inhomogeneity always
decreases entanglement, in contrast to the ferromagnetic case.
\begin{figure}\label{mar4}
\begin{center}
\epsfig{file=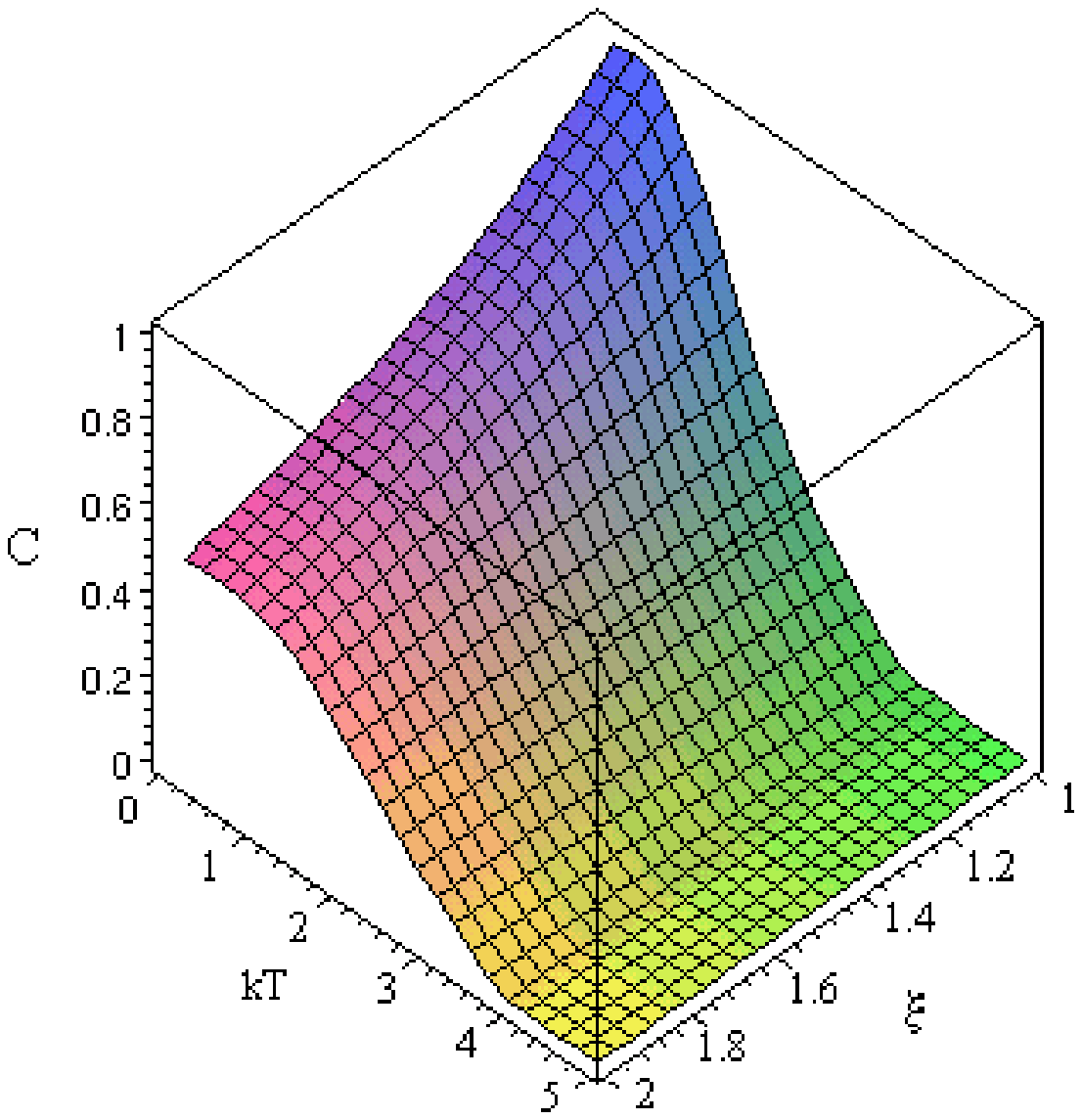,width=10cm}\caption{(color online)
Concurrence versus temperature and inhomogeneity in zero magnetic
field in the anti-ferromagnetic system.}\end{center}
\end{figure}

\begin{figure}\label{threshold}
\begin{center}
\epsfig{file=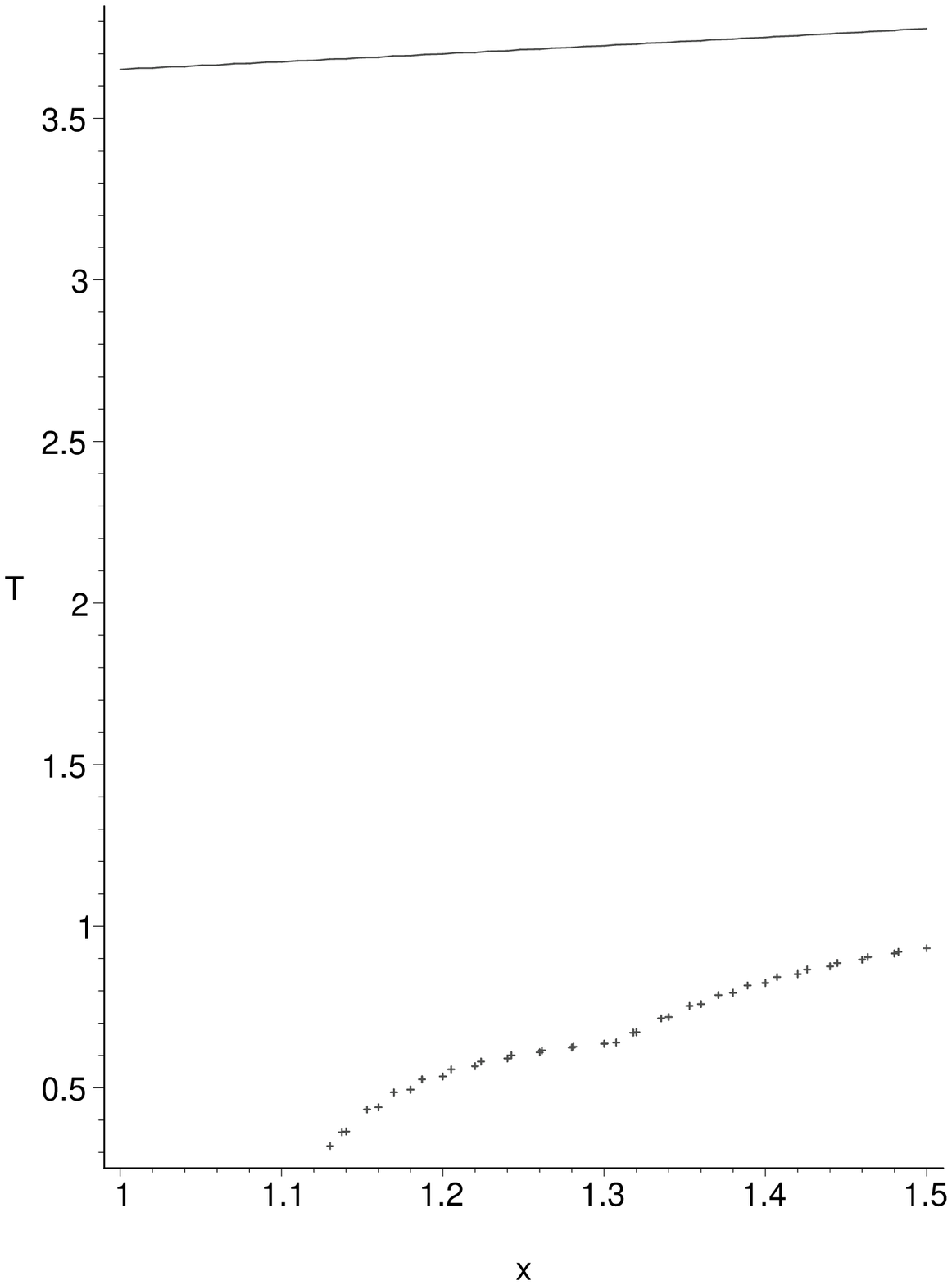,width=8 cm}\caption{The variation of
threshold temperature with inhomogeneity (denoted here as x) of
the magnetic field in the ferromagnetic (dotted line) and
anti-ferromagnetic (solid line) cases.}\end{center}
\end{figure}
\section{Discussion}
We have studied the effect of a inhomogeneous magnetic field on
the ground state entanglement and thermal entanglement of a two
spin system. We have shown that the effect of inhomogeneity is
most pronounced on ferromagnetic spins, i.e. spins coupled by
ferromagnetic interactions. At zero temperature an infinitesimal
magnetic field applied to the two spins in opposite directions
maximally entangles the two spins. It is as if we twist the two
spins into an entangled state. This effect also exists at higher
temperatures but to much less degree. When the coupling of the
spins is anti-ferromagnetic inhomogeneity can only have a
weakening effect on entanglement. Although we have derived our
results by studying a two spin systems, these results may also
hold true more or less on spin chains. A parameter like
$\xi:=\sqrt{1+\la b^2\ra}$, where $\la b^2\ra$ is the average of
inhomogeneity on all sites, i.e. $\la b^2\ra:=
\frac{1}{N}\sum_{i=1}^{N}(B_i-\la B\ra)^2$, may characterize the
influence of inhomogeneity on the entanglement of a spin chain.

\end{document}